\documentclass[review]{elsarticle}

\usepackage{hyperref}
\usepackage{amssymb}
\usepackage{graphicx}
\usepackage[ruled,vlined,linesnumbered]{algorithm2e}
\usepackage{float}
\usepackage{color}
\usepackage{soul}
\usepackage{amsmath}
\usepackage{caption}
\usepackage{makecell}
\usepackage{enumitem} 
\usepackage{makecell}
\usepackage{subcaption}
\journal{Journal: Internet of Things - Elsevier}

\begin{document} 

\begin{frontmatter}
\title{CyberLearning: Effectiveness Analysis of Machine Learning Security Modeling to Detect Cyber-Anomalies and Multi-Attacks}

\author{Iqbal H. Sarker$^{1,2*}$}
\address{$^1$Department of Computer Science and Engineering, \\ Chittagong University of Engineering \& Technology, \\ Chittagong-4349, Bangladesh. \\ $^2$Swinburne University of Technology, \\ Melbourne, VIC-3122, Australia. \\  
*Corresponding email: msarker@swin.edu.au \\
ORCID iD: https://orcid.org/0000-0003-1740-5517}

\begin{abstract}
Detecting cyber-anomalies and attacks are becoming a rising concern these days in the domain of cybersecurity. The knowledge of artificial intelligence, particularly, the \textit{machine learning} techniques can be used to tackle these issues. However, the effectiveness of a learning-based security model may vary depending on the security features and the data characteristics. In this paper, we present ``CyberLearning'', a machine learning-based cybersecurity modeling with \textit{correlated-feature selection}, and a \textit{comprehensive empirical analysis} on the effectiveness of various \textit{machine learning} based security models. In our CyberLearning modeling, we take into account a \textit{binary classification model} for detecting anomalies, and \textit{multi-class classification model} for various types of cyber-attacks. To build the security model, we first employ the popular ten machine learning classification techniques, such as naive Bayes, Logistic regression, Stochastic gradient descent, K-nearest neighbors, Support vector machine, Decision Tree, Random Forest, Adaptive Boosting, eXtreme Gradient Boosting, as well as Linear discriminant analysis. We then present the artificial neural network-based security model considering multiple hidden layers. The effectiveness of these learning-based security models is examined by conducting a range of experiments utilizing the two most popular security datasets, UNSW-NB15 and NSL-KDD. Overall, this paper aims to serve as a reference point for data-driven security modeling through our experimental analysis and findings in the context of cybersecurity.
\end{abstract}

\begin{keyword}
\texttt{cybersecurity; machine learning; deep learning; classification; feature selection; anomaly detection; cyber-attacks; security intelligence; cyber data analytics; intelligent systems.}
\end{keyword}

\end{frontmatter}


\section{Introduction}
In recent days, the demand for cybersecurity and protection against cyber-anomalies and various types of attacks, such as unauthorized access, denial-of-service (DoS), botnet, malware, or worms has been ever increasing. Such anomalies led to irreparable damage and financial losses in large-scale computer networks \cite{sarker2020intrudtree} \cite{moustafa2015unsw}. For example, one ransomware virus in May 2017 caused tremendous losses to many organizations and sectors, including banking, medical care, electricity, and universities, and caused a loss of 8 billion dollars \cite{qu2019survey}. In the domain of cybersecurity, such security breaches or intrusions have become the common issue these days while securing a cyber-system as well as an Internet of Things (IoT) system. Although various traditional methods, such as firewalls, encryption, etc., are designed to handle Internet-based cyber-attacks, an intelligent system that effectively detects such anomalies or attacks, is the key to tackle these issues. Thus, in this paper, we mainly focus on the knowledge of artificial intelligence, particularly, the applicability of \textit{machine learning security modeling}, which could be more effective due to its automated learning capabilities from the training security data.

Developing machine learning-based security models to analyze various cyber-attacks or anomalies, and eventually detect or predict the threats can be used for intelligent security services \cite{sarker2021ai}. Typically, the detection models could be for handling multiple associated cyber-attacks, i.e., ``multi-class" problem, or to detect anomalies, i.e., ``binary-class" problem. Several recent research, such as to detect botnet attack \cite{soe2020machine}, attack and anomaly detection analysis in IoT sensors in IoT site \cite{hasan2019attack}, classifying attacks to build an intrusion detection system \cite{raman2019efficient}, to detect the anomalous network connections and classifying the normal traffic and attack \cite{malik2018hybrid}, etc. have been done in the area. Although several machine learning techniques are used for different purposes, these are limited to analyze the variations in the significance of the security features, or to conduct the empirical analysis in a small range in terms of techniques used for security intelligence modeling. These are discussed briefly in Section \ref{Background}, and summarized in Table \ref{classifiers-summary}. Moreover, in case of unknown attacks, the abnormal behaviors that are considered as anomalies, which is different from the normal traffic, and the relevant model can be used in many security solutions \cite{sarker2020intrudtree} \cite{sarker2020cybersecurity}. Thus to classify the associated attacks in several well-known classes such as DoS, botnet, malware, worms, etc. as well as to classify anomalies for unknown attacks from the normal traffic is essential for intelligent modeling in the area of cybersecurity.

Different machine learning models by taking into account the above-mentioned issues may perform differently according to their learning capabilities from security data. The reason is that the effectiveness of a learning-based security model may vary depending on the significance of the associated \textit{security features} and the \textit{data characteristics}. In the real-world scenario, the cybersecurity issues might be involved with a huge number of security features, several known or unknown attack classes, or anomalies. Thus, an effective feature selection technique and a robust classification model usually consist of the construction of an intelligent intrusion detection system. Various types of machine learning techniques and their applicability in the area of cybersecurity, have been discussed briefly in Sarker et al. \cite{sarker2020cybersecurity}, however a detailed empirical analysis is needed by taking into account the above-mentioned issues to make an intelligent decision in the area. Therefore, we aim to present a comprehensive empirical analysis on the effectiveness of various \textit{machine learning} based security models by taking into account the issues, to make an intelligent decision in such diverse real-world scenarios in the area. 

To address the issues mentioned above, in this paper, we present ``CyberLearning'', a machine learning-based security modeling by taking into account the significance of the security features, and relevant experimental analysis. In our analysis, we take into account a \textit{binary classification model} for detecting anomalies, and \textit{multi-class classification model} for detecting various types of cyber-attacks, such as DoS, Backdoor, Worms, etc. In a binary-class classification model, the given security dataset is categorized into two classes, such as `normal' or `anomaly', whereas in a multi-class classification model, the given dataset is categorized into several attack classes, mentioned above. For modeling, we first employ the popular ten machine learning classification techniques, such as Naive Bayes (NB), Logistic regression (LR), Stochastic gradient descent (SGD), K-nearest neighbors (KNN), Support vector machine (SVM), Decision Tree (DT), Random Forest (RF), Adaptive Boosting (AdaBoost), eXtreme Gradient Boosting (XGBoost), Linear discriminant analysis (LDA), as well as Artificial Neural Network (ANN) based model, which is frequently used in deep learning \cite{sarker2021machine} \cite{sarker2021deep}. For selecting features, we take into account the feature correlation values, and then the resultant security model has been built based on the selected features considering both the model accuracy and simplicity or complexity. The main idea is that the learning-based model typically examines the behavior of the network utilizing the data, finding the security patterns for profiling the normal behavior, and thus detects the anomalies or associated attacks. The effectiveness of these learning-based security models is examined by conducting a range of experiments utilizing the two most popular security datasets, UNSW-NB15 \cite{moustafa2015unsw} and NSL-KDD \cite{tavallaee2009detailed}.

The contributions of this work can be summarized as follows. 

\begin{itemize}
	\item We first highlight the importance of \textit{security features} in a machine learning security modeling to detect cyber-anomalies and multi-attacks. Thus we adopt a correlated-feature selection approach to reduce the insignificant or irrelevant security features, which makes the security model lightweight and more applicable.
	
	\item We present a binary classification model for detecting \textit{cyber-anomalies} or unknown attacks, where the security model classifies the data into two classes, such as `normal' and `anomaly'. We also analyze the effectiveness of various popular machine learning classification models while detecting such anomalies.
	
	\item We present a multi-class classification model for detecting various \textit{cyber-attacks}, such as DoS, Backdoor, Worms, etc. where the security model classifies the data into these attack classes. We also analyze the effectiveness of various popular machine learning classification models while detecting such cyber-attacks.

	\item Finally, we conduct a range of experiments and present a \textit{comprehensive empirical analysis} on the effectiveness of various machine learning classification based security modeling for unknown test cases.
\end{itemize}  

The rest of the paper is organized as follows. Section \ref{Background} provides the background and related work of our study. In Section \ref{Methodology}, we present our machine learning-based security modeling by taking into account the significance of the security features. We evaluate the resultant security model and report the experimental results in Section \ref{Evaluation}. In Section \ref{Discussion}, several key findings of our analysis in the area are summarized. Finally, Section \ref{Conclusion} concludes this paper and highlights the future work.

\section{Background and Related Work}
\label{Background}
A number of research has been done in the area of cybersecurity with the capability of detecting cyber-anomalies and attacks or intrusions. In the cyber industry, both the signature-based intrusion detection system (SIDS) and anomaly-based intrusion detection systems (AIDS) are well-known for detecting and preventing cyber-attacks \cite{sarker2020cybersecurity}. SIDS is based on known signatures of the attacks \cite{seufert2007machine}. AIDS, on the other hand, has the benefit of identifying invisible threats over SIDS, including the ability to distinguish unknown or zero-day attacks \cite{alazab2012using} \cite{buczak2015survey}. Although association analysis is popular in the area of machine learning to build rule-based intelligent systems \cite{agrawal1994fast} \cite{sarker2020abc} \cite{sarker2019context}, it might not be effective due to its redundant generation and complexity with higher dimensions of security features while detecting anomalies or cyber-attacks. Thus, to achieve our goal, in this work, we primarily focus on machine learning classification models \cite{sarker2019classifications}, for security modeling because of their automated learning capabilities from the security data.

Several machine learning techniques have been used for various purposes. For instance, Li et al. \cite{li2012efficient} classify different types of attacks such as DoS, Probe or Scan, U2R, R2L, as well as regular traffic using SVM classifier using the most common KDD'99 cup dataset. Similarly, Amiri et al. \cite{amiri2011mutual}, Wagner et al. \cite{wagner2011machine}, Kotpalliwar et al. \cite{kotpalliwar2015classification}, Saxena et al. \cite{saxena2014intrusion}, Pervez et al. \cite{pervez2014feature}, Li et al. \cite{li2012efficient}, Shon et al. \cite{shon2005machine}, Kokila et al. \cite{kokila2014ddos}, and Raman et al. \cite{raman2019efficient} used SVM classifier in their studies for the purpose of detecting attacks. Several other classifiers are used to detect intrusions or attacks, in addition to the SVM classifier mentioned above. For example, a probability-based Bayesian network is used by Kruegel et al. \cite{kruegel2003bayesian} to identify events processing TCP/IP packets. Benferhat et al. have identified a DoS intrusion detector using the same Bayesian network in their research \cite{benferhat2008naive}. Similarly, Panda et al. \cite{panda2007network}, Koc et al. \cite{koc2012network} also use the naive Bayes classifier for detecting attacks in their systems.

Several studies \cite{bapat2018identifying} \cite{besharati2019lr} have been conducted to classify malicious traffic and intrusions using a logistic regression model. The KNN, an instance-based learning algorithm, is another common method of machine learning where the classification of a point is determined by that data point's k-nearest neighbors. Vishwakarma et al. \cite{vishwakarma2017intrusion}, Shapoorifard et al. \cite{shapoorifard2017intrusion}, Sharifi et al. \cite{sharifi2015intrusion} use KNN classification technique in their studies for the purpose of intrusion detection systems. Authors in \cite{kumar2011distributed} consider neural classifier, and in \cite{dainotti2009cascade} consider wavelet transform for anomaly detection particularly DoS attacks. A significant number of research in the domain of cybersecurity, such as Relan et al. \cite{relan2015implementation}, Rai et al. \cite{rai2016decision}, Ingre et al. \cite{ingre2017decision}, Malik et al. \cite{malik2018hybrid}, \cite{ingre2017decision}, Puthran et al. \cite{puthran2016intrusion}, Moon et al. \cite{moon2017dtb}, Balogun et al. \cite{balogun2015anomaly}, Sangkatsanee et al. \cite{sangkatsanee2011practical} use DT classification approach in their studies for the purpose of building intrusion detection systems. To detect anomalies and address loT cybersecurity threats in smart city, Alrashdi et al. \cite{alrashdi2019ad} use RF learning consisting of multiple decision trees in their binary classification model. Mazini et al. \cite{mazini2019anomaly} use AdaBoost approach with feature selection while building anomaly network-based intrusion detection system in their work.

A machine learning security model for detecting anomalies has been presented in \cite{sarker2020intrudtree}, which is effective in terms of prediction accuracy as well as reducing the feature dimensions based on the decision tree classification approach with feature selection. Recently, a machine learning-based botnet attack detection framework with sequential detection architecture has been presented in \cite{soe2020machine}, where ANN, DT, and NB classification techniques are used. Hasan et al. \cite{hasan2019attack} perform attack detection analysis in IoT sites, to develop a smart, secured, and reliable IoT based infrastructure. Although several machine learning techniques, such as SVM, DT, RF, LR, and ANN are used, the analysis is limited to a small number of security features for detecting different types of attacks. Moreover, the variations in the significance of the security features, which could be a crucial part while building an effective security model using machine learning techniques, are not addressed.

\begin{table*}[h]
	\tiny
	\centering
	\caption{A summary of machine learning based security models for detecting cyber-anomalies and attacks}
	\label{classifiers-summary}
	\begin{tabular}{|c|c|c|c|} 
		\hline
		\bf \makecell{Purposes} & \bf \makecell{Used Techniques} & \bf \makecell{Type} & \bf \makecell{References}\\  
		\hline
		
		\makecell{To detect IoT-Botnet Attack} & \makecell{ANN, DT, NB and \\ Feature selection} & Multiclass  &  \makecell{Soe et al. \cite{soe2020machine} \\ (2020)} \\     
		\hline
		
		\makecell{Classifying attacks to build an \\ efficient intrusion detection system} & \makecell{SVM and \\ Feature selection} & Multiclass &  \makecell{Raman et al. \cite{raman2019efficient} \\ (2019)} \\      
		\hline
		
		\makecell{To design a host-based intrusion \\ detection system} & \makecell{LR and \\ Feature selection} & Multiclass &  \makecell{Besharati et al. \cite{besharati2019lr} \\ (2019)} \\     
		\hline
		
		\makecell{To detect attacks in the \\ IoT environment} & \makecell{LR, SVM, DT, \\ RF, and ANN} & Multiclass &  \makecell{Hasan et al. \cite{hasan2019attack} \\ (2019)} \\      
		\hline
		
		\makecell{To build anomaly network-based \\ intrusion detection system} & \makecell{AdaBoost and \\, Feature selection} & Multiclass &  \makecell{Mazini et al. \cite{mazini2019anomaly} \\ (2019)} \\      
		\hline

		\makecell{Detecting attacks to establish an \\ efficient intrusion detection system} & \makecell{SVM and \\ Feature removal} & Multiclass &  \makecell{Li et al. \cite{li2012efficient} \\ (2012)} \\      
		\hline
		
		\makecell{To classify network events as normal \\ or attack events} & \makecell{NB and \\ Feature selection} & Multiclass &  \makecell{Koc et al. \cite{koc2012network} \\ (2012)} \\      
		\hline

		\makecell{To detect intrusion to the cloud system} & \makecell{KNN and \\ Feature selection} & Multiclass &  \makecell{Sharifi et al. \cite{sharifi2015intrusion} \\ (2015)} \\     
		\hline
		
		\makecell{To detect the anomalous network \\ connections and classifying normal and attack} & \makecell{DT \\ and pruning} & Binary  &  \makecell{Malik et al. \cite{malik2018hybrid} \\ (2018)} \\      
		\hline
		
		\makecell{To detect anomalies and address \\ loT cybersecurity threats in Smart City} & \makecell{RF learning} & Binary  &  \makecell{Alrashdi et al. \cite{alrashdi2019ad} \\ (2019)} \\      
		\hline
		
		\makecell{To detect anomalies in a network \\ and classifying normal and attack} & \makecell{DT and \\ Feature selection} & Binary  &  \makecell{Sarker et al. \cite{sarker2020intrudtree} \\ (2020)} \\    
		\hline
		
		\makecell{To introduce cybersecurity data science \\ highlighting cyber-anomalies and attacks} & \makecell{Overall machine \\ learning perspective} & --  &  \makecell{Sarker et al. \cite{sarker2020cybersecurity} \\ (2020)} \\    
		\hline
		
		\makecell{To detect cyber-anomalies and \\ multi-attacks} & \makecell{NB, LDA, KNN, XGBoost, \\ DT, RF, SVM, SGD, \\ AdaBoost, LR, ANN, and \\ Feature selection} & \makecell{Binary and \\ Multiclass}  &  \makecell{CyberLearning \\
			(our analysis)} \\      
		\hline
		
	\end{tabular}
\end{table*}

In the real-world scenario, the cybersecurity issues might be involved with a huge number of security features, and the effectiveness of a learning-based security model may vary depending on the significance of the associated security features and the data characteristics. Various types of machine learning techniques and their applicability in the area of cybersecurity, have been discussed in Sarker et al. \cite{sarker2020cybersecurity}, however a detailed empirical analysis is needed to make an intelligent decision in the area. Unlike the above approaches, in this paper, we present ``CyberLearning'', a machine learning-based cybersecurity modeling with correlated-feature selection according to their significance in modeling, and a comprehensive empirical analysis on the effectiveness of various machine learning-based security models. While building the security models, we take into account a binary classification model for detecting anomalies, and a multi-class classification model for detecting multi-attacks in the context of cybersecurity, to provide a comprehensive view to the readers in the area. In Table \ref{classifiers-summary}, we also summarize the most relevant machine learning-based security models within the scope of our study for a clear understanding for the readers.

\section{Materials and Methods}
\label{Methodology}
In this section, we present our security model of machine learning to detect cyber-anomalies and attacks. This involved several processing steps: exploring the security dataset, preparing raw data, determining the correlation and ranking of features, and constructing a security model. We address these steps briefly in the following section in order to achieve our goal.

\subsection{Exploring Security Dataset}
Usually, security datasets reflect a series of information records consisting of several security features and relevant details that can be used to construct a security model \cite{sarker2020cybersecurity} for detecting anomalies. Thus, to detect malicious activity or anomalies, it is important to understand the nature of raw cybersecurity data and the trends of security incidents. In this work, we use the most popular UNSW-NB15 \cite{moustafa2015unsw} and NSL-KDD \cite{tavallaee2009detailed} security datasets, to build the data-driven security model and the effectiveness analysis. 

\begin{table}[H]
	\centering
	\tiny
	\caption{UNSW-NB15 Dataset features with value type.}
	\label{feature-type}
	\begin{tabular}{cccc} 
		
		\textbf{Feature Name} &  \textbf{Value Type}  &  \textbf{Feature Name} & \textbf{Value Type} \\  
		
		$srcip$ &  Nominal    &   $sport$  &   Integer  \\
		$dstip$ &  Nominal & $dsport$  &  Integer  \\	
		$proto$ &  Nominal & $state$  &  Nominal  \\
		$dur$ &  Float & $sbytes$  &  Integer  \\
		$dbytes$ &  Integer & $sttl$  &  Integer  \\
		
		$dttl$ &  Integer & $sloss$  &  Integer  \\
		$dloss$ &  Integer & $service$  &  Nominal  \\
		$Sload$ &  Float & $Dload$  &  Float  \\
		$Spkts$ &  Integer & $Dpkts$  &  Integer  \\
		$swin$ &  Integer & $dwin$  &  Integer  \\
		
		$stcpb$ &  Integer & $dtcpb$  &  Integer  \\
		$smeansz$ &  Integer & $dmeansz$  &  Integer  \\
		$Sload$ &  Float & $Dload$  &  Float  \\
		$Spkts$ &  Integer & $Dpkts$  &  Integer  \\
		$swin$ &  Integer & $dwin$  &  Integer  \\
		
		$trans\_depth$ &  Integer & $res\_bdy\_len$  &  Integer  \\
		$Sjit$ &  Float & $Djit$  &  Float  \\
		$Stime$ &  Timestamp & $Ltime$  &  Timestamp \\
		$Sintpkt$ &  Float & $Dintpkt$  &  Float  \\
		$tcprtt$ &  Float & $synack$  &  Float  \\
		
		$ackdat$ &  Float & $is\_sm\_ips\_ports$  &  Binary \\
		$ct\_state\_ttl$ &  Integer & $ct\_flw\_http\_mthd$  & Integer \\
		$is\_ftp\_login$ &  Binary & $ct\_ftp\_cmd$  & Integer \\
		$ct\_srv\_src$ &  Integer & $ct\_srv\_dst$  & Integer \\
		$ct\_dst\_ltm$ &  Integer & $ct\_src\_ltm$  & Integer \\
		
		$ct\_src\_dport\_ltm$ &  Integer & $ct\_dst\_sport\_ltmltm$  & Integer \\
		$ct\_dst\_src\_ltm$ &  Integer &   &  \\
	\end{tabular}
\end{table}

Nine types of attacks, including Fuzzers, Study, Backdoors, DoS, Exploits, Generic, Reconnaissance, Shellcode, and Worms, are included in the UNSW-NB15\cite{moustafa2015unsw} dataset. It contains 257673 instances with the training and testing set and 45 features. On the other hand, NSL-KDD \cite{tavallaee2009detailed} dataset contains the Denial of Service Attack (DoS), User to Root Attack (U2R), Remote to Local Attack (R2L), and Probing Attack. The raw data source consists of 494020 instances with 41 security features that are taken into account in our experimental analysis. The features can be in various types in a dataset. For instance, in Table \ref{feature-type}, we show the security features of the UNSW-NB15 dataset, where the features are not identical. Thus effectively analyzing these features and building a security model for detecting the anomalies and multi-attacks mentioned above, is the key in our analysis.

\subsection{Security Data Pre-Processing}
Data preparation includes anomaly and attacks, feature encoding, and scaling according to the characteristics of the given dataset.

\begin{itemize}
	\item \textit{Anomaly and Attacks:} As mentioned earlier, the dataset UNSW-NB15 \cite{moustafa2015unsw} contains nine types of attacks. These are known as anomalies in this dataset and are used in a binary classification model, while all these separate attacks are used in a multi-class classification model that is taken into account in our analysis. Similarly, the four types of attacks such as DoS, U2R, R2L, and Probing, are known as anomalies in NSL-KDD \cite{tavallaee2009detailed} dataset and are used in the corresponding classification model.
	
	\item \textit{Feature encoding:} As shown in Table \ref{feature-type}, the dataset UNSW-NB15 \cite{moustafa2015unsw} contains several feature types such as the nominal, integer, float, timestamp, and binary values. Thus, to fit the data to the security model, we first convert all the nominal valued features into vectors. Although, ``One Hot Encoding'' is a popular technique, we use ``Label Encoding'' in this work. The reason is that, in one hot encoding technique, a significant number of feature dimensions increase \cite{sarker2020intrudtree}. The label-encoding technique, on the other hand, transforms the feature values directly into precise numeric values that can be used to fit a classification model for machine learning. Similarly, the features in NSL-KDD \cite{tavallaee2009detailed} dataset are encoded to build the resultant security model.
	
	\begin{figure*}
		\centering
		\begin{minipage}{.5\textwidth}
			\centering
			\includegraphics[width=.95\linewidth]{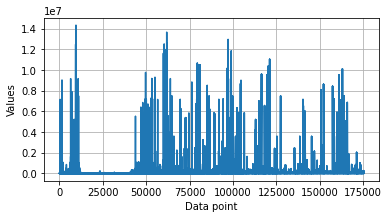}
			\captionof{figure}{Secuity feature $`sbyte'$}
			\label{fig:data-distribution-sbyte}
		\end{minipage}%
		\begin{minipage}{.5\textwidth}
			\centering
			\includegraphics[width= .95\linewidth]{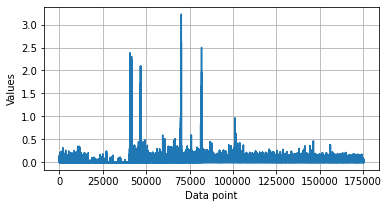}
			\captionof{figure}{Secuity feature $`synack'$}
			\label{fig:data-distribution-synack}
		\end{minipage}
	\end{figure*}
	
	\item \textit{Feature scaling:} Feature scaling is also known as data normalization in the task of data pre-processing. All the security features in a dataset may not identical in terms of data distribution, and vary from feature to feature. For instance, Figure \ref{fig:data-distribution-sbyte} and Figure \ref{fig:data-distribution-synack} show the data distribution for two different features, $sbyte$ and $synack$ respectively in the dataset UNSW-NB15 \cite{moustafa2015unsw}. According to Figure \ref{fig:data-distribution-sbyte} and Figure \ref{fig:data-distribution-synack}, for some data points, the value is very low while for some data points, it is much higher. Thus, we use Standard Scaler, a data scaling method that is used to normalize the range of the feature values with the mean value = 0 and standard deviation = 1.
	
	\item \textit{Data Splitting:} As we aim to build learning-based security modeling, data splitting can be considered as an important part. The reason is that a good security model may be based on bad data splitting. Thus, for building a fair model and evaluation, we first consider the data from data sources as input data and split them using a $k$ fold cross-validation technique \cite{han2011data}. According to $k$ fold cross-validation technique, we first randomly partition the input data mentioned above into $k$ mutually exclusive subsets or ``folds'', $d_1,d_2,...,d_k$. Each fold has an approximately equal size of data instances. The model needs $k$ iteration to complete the overall process. Thus, in each iteration $i$, we use all the data instances of all folds except $d_i$ as the training dataset that can be used to build the resultant security model. For evaluation purpose $d_i$ is used as the testing dataset in each iteration $i$. Eventually, the average result is taken into account as the outcome of the model.
\end{itemize}

\subsection{Modeling Techniques}
In our CyberLearning modeling, we take into account the impact of security features while building the security model. In the following, we present how we rank the features for selection, and various machine learning algorithms that are employed to build the model, and effectiveness analysis within the scope of our study.

\subsubsection{Feature Ranking and Selection}
Feature selection in the cybersecurity domain can provide a better understanding of the security data, a way of simplifying the security model by reducing the computational cost or model complexity, as well as providing significant outcomes in a machine learning-based model. Security dataset may contain data with high dimensions, and some of them may be highly correlated to anomalies or attacks, while some have less correlation or no correlation at all. Thus, in order to create a machine learning classification-based security model, all the security features in a given dataset may not contain significant details. In addition, due to the over-fitting issue \cite{sarker2020intrudtree} \cite{sneha2019analysis}, further processing with all the security features could provide poor results. Thus, security feature selection is required not only to reduce the computational cost but also to create a more efficient security model with a higher accuracy rate. Thus, security feature selection is considered as a method that can be used to filter those features that are less significant, redundant, or have no impact on modeling, from the given security dataset. 

To achieve this goal, we first calculate the correlation of the security features, known as the Pearson correlation coefficient, and rank them accordingly. The correlation-based feature selection is based on the following hypothesis: ``Good feature subsets contain features highly correlated with the target class, yet uncorrelated or less correlated to each other''. If $X$ and $Y$ represent two random contextual variables, then the correlation coefficient between $X$ and $Y$ is defined as \cite{han2011data} -

\begin{equation} 
\label{correlation}
r (X,Y) = \frac{\sum_{i=1}^{n} (X_i - \bar X) (Y_i - \bar Y)}{\sqrt{\sum_{i=1}^{n} (X_i - \bar X) ^2} \sqrt{\sum_{i=1}^{n} (Y_i - \bar Y) ^2}}
\end{equation}

In the field of statistics, the formula Equ. \ref{correlation} is often used to determine how strong that relationship is between those two variables $X$ and $Y$. In our security modeling, the higher the value, the more significant the security feature for building the resultant learning-based security model. For instance, a value of $1$ (max) means that the outcome of the learning-based security model is directly associated with that security feature, and $0$ (min) means that the output of the model does not depend on that security feature at all. Thus, in the scope of our analysis, we calculate the correlation coefficient values of each security feature in both our binary classification modeling for detecting anomalies and multi-class classification modeling for detecting various types of attacks.

\subsection{Machine Learning Algorithms and Parameters}
In this section, we present how various machine learning classification techniques as well as ANN-based modeling with multiple hidden layers, are used in our security modeling.

\subsubsection{Naive Bayes (NB)}
Naïve Bayesian (NB) \cite{john1995estimating} is one of the common classification techniques for machine learning that is often used in the field of machine learning and data science. This is based on Bayes's theorem that describes the probability of a given feature, according to the prior knowledge of situations related to that feature. Let, $X = \{x_1, x_2,..., x_n\}$ is a security feature vector of size $n$, and $c$ is a class variable that represents the cyber-attacks or anomalies. Thus, it calculates the probability $(P)$ using the following equation \cite{han2011data}:

\begin{equation}
P(c|X) =\frac{P(X|c)P(c)}{P(X)}
\end{equation}
\begin{equation}
P(c|x_1, x_2,..., x_n) = \frac{P(x_1|c)P(x_2|c)... P(x_n|c) P(c)} {P(x_1)P(x_2)...P(x_n)}
\end{equation}

To build a security model, we use the Gaussian Naive Bayes classifier \cite{pedregosa2011scikit} assuming all the security features are following a Gaussian distribution i.e, normal distribution. The prior probabilities of the classes in our security modeling are adjusted according to the data. The portion of the largest variance of all security features is added to the variances for calculation stability or smoothing.

\subsubsection{Linear Discriminant Analysis (LDA)}
In machine learning, Linear Discriminant Analysis (LDA) \cite{han2011data} is another probability-based method to find a linear combination of security features that separates the anomaly or attack classes. This method is also known as a generalization of Fisher's linear discriminant, that projects a given security dataset onto a lower-dimensional space, i.e., dimensionality reduction that minimizes the model complexity or reduce the computational costs of the resultant security model. Consequently, it has the capability for good class-separability to avoid the problem of overfitting. Thus, the resulting combination mentioned above can be used as a linear classifier or, more specifically, for dimensionality reduction of security features before performing the tasks of anomaly or attack classification. The standard LDA model typically fits a Gaussian density to each class such as `anomaly' or `normal' or various types of attacks, assuming that all classes share the same covariance matrix \cite{pedregosa2011scikit}. For modeling, the LDA approach also uses Bayes' theorem mentioned above to estimate probabilities and to make predictions of the class anomaly or various types of cyber-attacks based upon the probability that a new input dataset belongs to each anomaly or attack class. The class which has the highest probability is considered the output anomaly or attack class, and then the LDA makes a prediction. In our security modeling, we use $singular \; value \; decomposition$ as a solver method with no shrinkage to get the outcome. The prior probabilities of the classes in our security modeling are inferred from the given security data.

\subsubsection{K-nearest Neighbor (KNN)}	
K-nearest neighbors (KNN) \cite{aha1991instance}, also known as a lazy learning algorithm, is an instance learning or non-generalizing learning. Instead of using all data instances during classification, this approach does not have a specialized training process for constructing a model. Based on a 'feature similarity' scale, it classifies new test cases, considering a distance function, such as $Minkowski$, $Euclidean$, $Manhattan$ distance etc \cite{han2011data}. Let, two variables $X$ and $Y$, then the $Minkowski \; distance$ between these two variables is defined as $\left(\sum_i |X_i-Y_i|^p \right)^{1/p},\ \text{where}\ p\ge1$. It can behave differently depending on $p$ values, such as $p = 1$ and $p = 2$ represent Manhattan and Euclidean distance respectively.

\begin{equation}
\label{Euclidean}
d\left( X,Y\right)   = \sqrt {\sum _{i=1}^{n}  \left( X_{i}-Y_{i}\right)^2 } 
\end{equation}

In our security modeling, we take into account the most popular Euclidean distance considering $p = 2$ \cite{pedregosa2011scikit}, and can be defined as Equ \ref{Euclidean}. The number of neighbors indicating as $k$ values is another key parameter in a KNN based security modeling. Thus, we take into account $k=5$, as the number of neighbors, and uniform weights, where all points in each neighborhood are weighted equally in our security modeling.

\subsubsection{Decision Tree (DT)}
Decision tree (DT) \cite{quinlan1993} is a well-known classification framework for machine learning, which is commonly used in various fields of use. A decision tree is a method of non-parametric supervised learning that breaks down a given security dataset into smaller subsets and incrementally generates a related branch of the tree. For splitting, the most popular criteria are ``gini'' for the Gini impurity and ``entropy'' for the information gain, which can be expressed mathematically as \cite{pedregosa2011scikit}. 

\begin{equation}
Entropy: H(x) = -\sum_{i=1}^n p(x_i) \log_2 p(x_i)
\end{equation}

\begin{equation}
\label{Gini}
Gini (E) = 1 - \sum_{i=1}^{c}{p_i}^2
\end{equation}

Where $p_i$ denotes the probability of an element being classified for a distinct anomaly or attack class. To build a decision tree based security model, we use ``Gini Index'' that is determined by deducting the sum of squared of probabilities of each anomaly or attack class from $1$ that can be expressed as Equ. \ref{Gini}. While generating the tree considering both the anomaly or attack classes, nodes are taken into account to expand until all leaves are pure or until all leaves contain less than two sample instances.

\subsubsection{Random Forest (RF)}
In the field of machine learning and data science, the random forest (RF) \cite{breiman2001random} is well known as an ensemble classification technique that is used in different application areas. This consists of multiple decision trees, where a decision tree classifier discussed above is used as a single tree in the forest model. This combines the bootstrap aggregation (bagging) \cite{breiman1996bagging} with the random selection of features \cite{amit1997shape} to create a collection of controlled variance decision trees. The majority voting of the generated decision trees in a forest model is used to measure the outcome. To build a random forest security model, we generate $N=100$ decision trees in the forest, where the quality of a split in a tree is measured by `Gini', defined earlier in Equ. \ref{Gini}. 

\subsubsection{Support Vector Machine (SVM)}
In machine learning, support vector machine (SVM) \cite{han2011data} is another popular classification technique. This technique is based on a hyperplane between the data space, which best divides the security dataset into two classes, such as `anomaly' or `normal' and can behave differently based on the mathematical functions known as the kernel that can be different types such as linear, nonlinear, polynomial, radial basis function (RBF), sigmoid, etc. To build a security model, we use the RBF kernel \cite{keerthi2001improvements}, also known as the Gaussian kernel, considering no prior knowledge about the given security data. The RBF kernel is mathematically defined as -

\begin{equation}
\label{RBF-kernel}
k(x, y) = exp(-\lambda||x-y||^2) 
\end{equation}

where $\lambda$ is a parameter that sets the ``spread'' of the kernel. Based on this RBF kernel function defined in Equ. \ref{RBF-kernel}, this technique manipulates the given security data accordingly to achieve the goal. Overall, it works in two stages, including the identification of the optimal hyperplane in the data space and then the mapping of the security data instances according to the hyperplane's defined decision boundaries. Moreover, we use $C=1.0$ (regularization parameter), considering the trade-off between achieving a low training error, and a low testing error in a SVM based security model.

\subsubsection{Logistic Regression (LR)}
Another common probabilistic dependent statistical model used to solve the classification problems in machine learning is Logistic Regression (LR) \cite{le1992ridge}. Typically, logistic regression calculates the probabilities using a logistic equation, which is often referred to as the mathematically defined sigmoid function -

\begin{equation}
\label{Logistic}
g(z) = \frac{1}{1 + exp(-z)}
\end{equation}

While building LR based security modeling, we use $L_2$ regularization, i.e., $Ridge$ regression that adds squared magnitude of coefficient as penalty term to the loss function. The ``C'' is similar to the SVM model. We also use Scikit-learn solver ``lbfgs'' \cite{pedregosa2011scikit}, which stands for Limited-memory Broyden–Fletcher–Goldfarb–Shanno, to build the security model.

\subsubsection{Adaptive Boosting (AdaBoost)}
Boosting, a machine-learning algorithm, can be used for classification that is able to reduce bias and variance from the dataset. Boosting helps to convert weak learners to strong ones. Adaptive Boosting (AdaBoost) is such an algorithm formulated by Yoav Freund et al. \cite{freund1996experiments}. In that sense, AdaBoost is called an adaptive classifier by significantly enhancing the efficiency of the classifier, but in some instances, it can trigger overfits. For noisy data and outliers, AdaBoost is sensitive. We use a decision tree classifier with maximum depth ($max\_depth=1$) as a base estimator. The maximum number of estimators is taken into account as $50$ at which boosting is terminated.

\subsubsection{Extreme Gradient Boosting (XGBoost)}
Gradient Boosting is another ensemble learning algorithm, similar to the Random Forests discussed above, that creates a final model based on a set of individual models. Similar to how neural networks use gradient descent to optimize weights, the gradient is used to minimize the loss function. XGBoost stands for Extreme Gradient Boosting, which is known as a special Gradient Boosting method that takes into account more accurate approximations to find the best model. It computes second-order gradients of the loss function to minimize the loss and advanced regularization (L1 \& L2), which reduces overfitting and improves model generalization. We employ scikit-learn \cite{pedregosa2011scikit} API compatible class while building a security model based on XGBClassifier in our analysis.

\subsubsection{Stochastic Gradient Descent (SGD)}	
Stochastic gradient descent (SGD) \cite{han2011data} is an iterative method for optimizing an objective function with suitable smoothness properties, where the word `stochastic' means a system or a process that is linked with a random probability. A gradient is the slope of a function that calculates a variable's degree of change in response to another variable's changes. Gradient Descent is mathematically a convex function whose output is a partial derivative of a set of its input parameters. Let, $\alpha$ is the learning rate, and $J_i$ is the cost of $i^{th}$ training example, then Equ. \ref{SGD} represents the weight update process for the stochastic gradient descent at $j^{th}$ iteration.   

\begin{equation}
\label{SGD}
w_j \ :=\ w_j - \alpha \ \frac{\partial J_i}{\partial w_j}
\end{equation}

The greater the gradient, the steeper the slope. While building the security model, we use a loss function $`hinge'$, which gives a linear SVM. Moreover, we use $L_2$ regularization similar to logistic regression and a constant $alpha = 0.0001$ that multiplies the regularization term while building the security model.

\subsubsection{Artificial Neural Network (ANN)}
Artificial Neural Network (ANN) is also a machine learning technique and used typically in deep learning modeling, which is comprised of a network of artificial neurons or nodes \cite{han2011data}. In this work, we build a feed-forward ANN-based deep learning security model consisting of an input layer with the selected security features, three hidden layers with 128 neurons, and an output layer with one neuron for binary classification, or the equal number of classes for multi-class classification task. We also use dropout in each layer to simplify the security model and compile the neural network model with Adam optimizer \cite{geron2019hands}. 

\begin{equation}
\label{ReLU}
ReLU: f(x) = max(0, x)
\end{equation}
\begin{equation}
\label{Softmax}
Softmax:  f(y_k) = \frac{\exp(\phi_k)}{\sum^{c}_j \exp(\phi_j)}
\end{equation}
\begin{equation}
\label{Sigmoid}
Sigmoid: f(z) = \frac{1}{1+e^{-z}}
\end{equation}
\begin{equation}
\label{Loss}
Loss = 
\begin{cases}
-{(y\log(p) + (1 - y)\log(1 - p))} & \texttt{for } binary \\
-\sum_{c=1}^My_{o,c}\log(p_{o,c}) & \texttt{for } multiclass \\
\end{cases}
\end{equation}

We use 100 epochs with a batch size of 128 when training the security network. We often use a small value of 0.001 as the learning rate, as it enables the global minimum to be reached by the security network model. We use the Rectified Linear Unit (ReLU) described in Equ. with regard to the activation function. \ref{ReLU}, which addresses the problem of the vanishing gradient, as well as helps the model to learn faster. However, we use the Softmax activation function defined in Equ. \ref{Softmax} for multi-class attack detection and the Sigmoid or Logistic activation function defined in Equ. \ref{Sigmoid} for binary classification as it exists between (0 to 1) in the output layer. To adjust the weights of the model, we use the Cross-Entropy loss function, defined in Equ. \ref{Loss}, where $M$ represents the number of attack classes $c$, $y$ represents binary indicator, and $p$ represents probability observation $o$. The popular Backpropagation technique \cite{han2011data} is used to adjust the connection weights between neurons of the security model during learning.  

\section{Experimental Results and Analysis}
\label{Evaluation}	
In this section, we aim to briefly analyze and report the experimental results of machine learning-based security modeling as well as artificial neural network-based model utilizing the security datasets. For this, we first set up our experiments highlighting several questions to evaluate our security model, and then briefly discuss the experimental results and findings in various dimensions related to our analysis of cyber-anomalies and multi-attacks detection.

\subsection{Experimental Setup}
To evaluate our CyberLearning model, we aim to answer the following questions:

\begin{itemize}
	\item Question 1: Does the impact of the security features vary from feature to feature while building a machine learning-based security model? 
	
	\item Question 2: How effective is the machine-learning-based security model for detecting cyber-anomalies considering binary classification?
	
	\item Question 3: How effective is the machine-learning-based security model for detecting multi-attacks considering multi-class classification?
	
	\item Question 4: How effective the artificial neural network-based security model for detecting the anomalies and multi-class attacks?
\end{itemize}

To answer these questions related to our CyberLearning analysis, we have conducted a range of experiments on security datasets consisting of the anomalies and multi-attacks discussed in the earlier section. We have implemented all these methods in Python programming language using Scikit-learn \cite{pedregosa2011scikit}, Tensorflow, and Keras \cite{geron2019hands}, and executed them on Google Colab \cite{GoogleColab}. In the following subsections, we first define the evaluation metrics that are taken into account in our experimental evaluation.

\subsection{Evaluation Metric}
To measure the effectiveness of our CyberLearning model, we compute the outcome results in terms of precision, recall, F-score, as well as model accuracy in percentage. For this, we first calculate the true positive rate (TP), true negative rate (TN), false positive rate (FP), and false-negative rate (FN) that are defined as below \cite{han2011data} -

\begin{itemize}
	\item TP (true positive): An outcome where the security model correctly detects or classifies the positive class of anomaly or attacks.
	\item TN (true negative): An outcome where the security model correctly detects or classifies the negative class of anomaly or attacks.
	\item FP (false positive): An outcome where the security model incorrectly detects or classifies the positive class of anomaly or attacks.
	\item FN (false negative): An outcome where the security model incorrectly detects or classifies the negative class of anomaly or attacks.
\end{itemize}

Based on these definitions of TP, TN, FP, and FN, we can compute the precision, recall, F-score, accuracy as below \cite{han2011data} -

\begin{equation}
Precision = \frac{TP}{TP + FP}
\end{equation}

\begin{equation}
Recall = \frac{TP}{TP + FN}
\end{equation}

\begin{equation}
F1-score = 2 * \frac{Precision * Recall}{Precision + Recall}
\end{equation}

\begin{equation}
Accuracy = \frac{TP + TN}{TP + TN + FP + FN}
\end{equation}

In the area of machine learning and data science, these metrics are well-known and widely used to measure the effectiveness of a model \cite{han2011data} \cite{sarker2019classifications}. The greater the value the effective the security model is. In the following subsection, we discuss the experimental results briefly and analyze the model effectiveness considering these metrics.

\subsection{Impact of Security Features and Ranking}
To answer the first question mentioned above, in this experiment, we calculate and show the impact of each feature based on their correlation values. Table \ref{feature-score} shows the calculated correlation scores of all the 42 security features utilizing the given security dataset UNSW-NB15. The results are shown in a descending order for detecting anomalies considering binary classification, where the values are arranged from the largest to the smallest number. If we observe Table \ref{feature-score}, we see that the calculated scores of all features are not identical in a given dataset, and may vary from feature-to-feature according to their impact on the target anomaly and attack classes. 

\begin{table}[H]
	\centering
	\tiny
	\caption{The ranking of the security features with corresponding correlation scores for detecting anomalies utilizing the dataset UNSW-NB15.}
	\label{feature-score}
	\begin{tabular}{cccccc} 
		
		\textbf{Rank} & \textbf{Feature} &  \textbf{Score} & \textbf{Rank} & \textbf{Feature} &  \textbf{Score} \\  
		
		01 & $sttl$ &  0.624082  & 22 & $dloss$ &  0.075961  \\ 		
		02 & $ct\_state\_ttl$  &   0.476559 & 23   &   $service$  &   0.073552  \\			
		03 & $state$ &  0.462972  & 24 & $dbytes$ &   0.060403  \\ 			
		04 & $ct\_dst\_sport\_ltm$  &  0.371672  & 25  &   $djit$  &  0.048819  \\
		05 & $swin$  &   0.364877  & 26 & $synack$ & 0.043250    \\   	
		
		06 & $dload$ &  0.352169 &	27 & $spkts$ &  0.043040  \\	
		07 &  $dwin$  &  0.339166  &	28   &   $dinpkt$  &   0.030136  \\		
		08 & $rate$ &  0.335883 &	29 & $dur$ &  0.029096  \\ 	
		09 & $ct\_src\_dport\_ltm$  &  0.318518  & 30  &   $smean$  &  0.028372  \\
		10 & $ct\_dst\_src\_ltm$  &   0.299609  & 	31 & $tcprtt$ & 0.024668    \\   	
		11 & $dmean$  &   0.295173 &  32 & $sbytes$ &  0.019376  \\			
		12   & $stcpb$  &   0.266585  &	 	33   &   $dttl$  &   0.019369  \\			
		13 & $dtcpb$ &   0.263543  & 34 & $response\_body\_len$ &  0.018930  \\ 		
		14  &   $ct\_src\_ltm$  &   0.252498  &		35  &   $sjit$  &  0.016436  \\	
		15  &   $ct\_srv\_dst$  &   0.247812  & 	36 & $ct\_flw\_http\_mthd$ & 0.012237    \\ 
		
		16  &   $ct\_srv\_src$  &  0.246596  &	37 & $ct\_ftp\_cmd$ &  0.009092  \\		
		17  &   $ct\_dst\_ltm$  &  0.240776  &	38   &   $is\_ftp\_login$  &   0.008762  \\		
		18  &   $sload$  &   0.165249  &		39 & $proto$ &  0.008023  \\ 	
		19  &   $is\_sm\_ips\_ports$  &   0.160126  & 40  &   $trans\_depth$  &  0.002246  \\	
		20  &   $sinpkt$  &   0.155454  & 	41  &   $sloss$  &  0.001828  \\
		
		21 & $dpkts$ & 0.097394    &  	42  &   $ackdat$  &  0.000817  \\			
	\end{tabular}
\end{table}

According to Table \ref{feature-score}, the feature $sttl$ has the highest score of $0.624082$ and thus selected as the top-ranked feature, whereas another feature $ackdat$ has a lower score of $0.000817$ that is closer to the value $0$ for this dataset, and thus selected as the last ranked feature. These correlation scores may be different for another dataset depending on their features and classes. The higher the correlation value, the more significant the feature in a security model. Thus, based on the scores, we can conclude that all the features in a given security dataset might not have a similar impact to build a data-driven security model.

\subsection{Effectiveness Analysis for Detecting Cyber-Anomalies}
To show the effectiveness of the security models based on machine learning classifiers, Table \ref{UNSW15-features-comparison-anomaly} shows the effectiveness comparison results in terms of accuracy (\%) for different machine learning classifier based anomaly detection models considering binary classification. The results in Table \ref{UNSW15-features-comparison-anomaly} are shown by varying the number of selected features such as 42, 31, 24, and 17 utilizing the dataset UNSW-NB15. These are selected according to their correlation scores and ranking, shown in Table \ref{feature-score} considering a particular threshold. If we observe the results in Table \ref{UNSW15-features-comparison-anomaly}, we can see that various machine learning security models have an impact on the number of selected features. In general, higher accuracy results considering a minimum number of top-ranked features represent the effectiveness of the security models, in terms of both the detection outcome and model complexity or simplicity. For instance, the NB security model gives higher accuracy (85\%) when the top 24 features are selected to build the model. Similarly, RF and SVM security models also give higher accuracy (95\%) and (92\%), when the top 24 features are selected to build the corresponding models. Some models such as LDA, AdaBoost, SGD, and LR show their significant results considering all the 42 features, while some models such as KNN, XGBoost show their significant results considering only the top 17 selected features. In addition to RF (accuracy 95\%), DT (accuracy 94\%), and XGBoost (accuracy 93\%) also give significant results for detecting anomalies. 

\begin{table}[H]
	\centering
	\tiny
	\caption{Effectiveness comparison results in terms of accuracy (\%) for different machine learning classifier based anomaly detection models utilizing the dataset UNSW-NB15.}
	\label{UNSW15-features-comparison-anomaly}
	\begin{tabular}{ccccc} 
		
		\textbf{Model} & \textbf{Features (42)} &  \textbf{Features (31)}  &  \textbf{Features (24)}  &  \textbf{Features (17)} \\  
		
		NB &	82 &	83 &	\textbf{85} &	75 \\
		LDA &	\textbf{89} &	87 &	87 &	84 \\
		KNN	& 92 &	92 &	92 &	\textbf{92} \\
		XGBoost &	93 &	93 &	93 &	\textbf{93} \\
		DT	& 94 &	\textbf{94} &	93 &	92 \\
		RF	& 95 &	95 &	\textbf{95} &	94 \\
		SVM	& 92 &	92 &	\textbf{92} &	91 \\
		AdaBoost & 	\textbf{93} &	92 &	92 &	92 \\
		SGD	& \textbf{89} &	88 &	88 &	86 \\
		LR	& \textbf{90} &	88 &	87 &	84 \\
	\end{tabular}
\end{table}

\begin{figure*}[htbp!]
	\centering
	\begin{subfigure}[b]{.49\textwidth}
		\includegraphics[width=\textwidth]{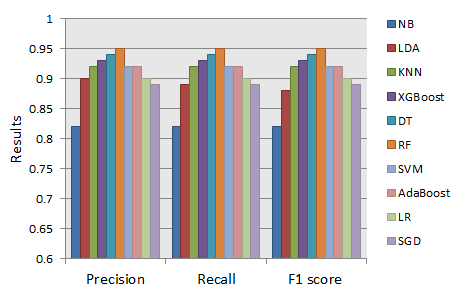}
		\caption{Anomaly detection with all 42 features}
		\label{fig:UNSW15-42-features-anomaly}
	\end{subfigure}
	\begin{subfigure}[b]{.49\textwidth}
		\includegraphics[width=\textwidth]{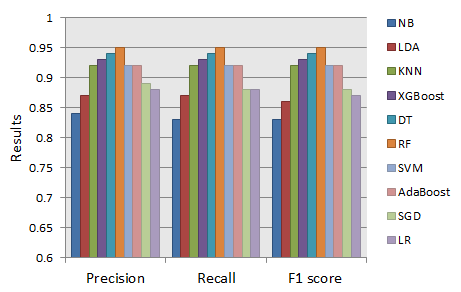}
		\caption{Anomaly detection with top 31 features}
		\label{fig:UNSW15-31-features-anomaly}
	\end{subfigure}
	\begin{subfigure}[b]{.49\textwidth}
		\includegraphics[width=\textwidth]{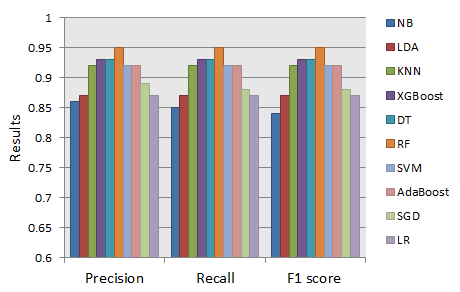}
		\caption{Anomaly detection with top 24 features}
		\label{fig:UNSW15-24-features-anomaly}
	\end{subfigure}
	\begin{subfigure}[b]{.49\textwidth}
		\includegraphics[width=\textwidth]{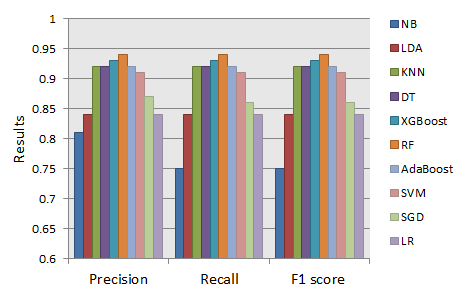}
		\caption{Anomaly detection with top 17 features}
		\label{fig:UNSW15-17-features-anomaly}
	\end{subfigure}
	\caption{Effectiveness comparison results in terms of precision, recall, and F1 score for different machine learning classifier based anomaly detection models utilizing the dataset UNSW-NB15.}
	\label{fig:UNSW15-features-comparison-anomaly}
\end{figure*}

In addition to Table \ref{UNSW15-features-comparison-anomaly}, Figure \ref{fig:UNSW15-features-comparison-anomaly} also shows the relative comparison of various security models based on machine learning classifiers for detecting anomalies. The comparative results are shown in terms of precision, recall, and F1 score for different numbers of top-ranked selected features such as 42, 31, 24, and 17 utilizing the dataset UNSW-NB15. For each security model, we use the same train and testing data to calculate these metrics for fair evaluation.

If we observe Figure \ref{fig:UNSW15-features-comparison-anomaly}, we find that tree-based classification models give higher prediction results than other security models, in terms of precision, recall, and F1 score, while applying on cybersecurity data consisting of various security features. In particular, the RF (Random Forest) based security model generating multiple decision trees gives the prediction results with the highest values of accuracy, recall, and F1 score for different number of features, shown in Figure \ref{fig:UNSW15-features-comparison-anomaly}. The interesting finding is that the RF model gives similar results with the features of 42, 31, and 24, and a comparatively lower result with feature 17. The reason for decreasing the result is that it losses significant information while reducing the features. Thus, the RF model with the top 24 security features is taken into account as an effective model considering both the accuracy and model complexity. Overall, based on the selected security features, we can conclude that the RF model gives better results in detecting cyber anomalies. The explanation is that the random forest model produces a collection of logical rules based on the chosen security features that take into account multiple decision trees created in the forest, and offers an outcome based on the majority vote of those trees.

\begin{table}[H]
	\centering
	\tiny
	\caption{Effectiveness comparison results in terms of accuracy (\%), precision, recall, and F1 score for different machine learning classifier based anomaly detection models utilizing the dataset NSL-KDD.}
	\label{NSL-KDD-features-comparison-anomaly}
	\begin{tabular}{ccccc} 
		
		\textbf{Model} & \textbf{Accuracy (\%)} &  \textbf{Precision}  &  \textbf{Recall}  &  \textbf{F1 Score} \\  
		
		NB &	98 &	0.98 &	0.98 &	0.98 \\
		LDA &	99 &	0.99 &	0.99 &	0.99 \\
		KNN	& 99 &	0.99 &	0.99 &	0.99 \\
		XGBoost &	99 &	0.99 &	0.99 &	0.99 \\
		DT	& 99 &	0.99 &	0.99 &	0.99 \\
		RF	& 99 &	0.99 &	0.99 &	0.99 \\
		SVM	& 99 &	0.99 &	0.99 &	0.99 \\
		AdaBoost & 	98 &	0.98 &	0.98 &	0.98 \\
		SGD	& 99 &	0.99 &	0.99 &	0.99 \\
		LR	& 98 &	0.98 &	0.98 &	0.98 \\
	\end{tabular}
\end{table}

In Table \ref{NSL-KDD-features-comparison-anomaly}, we also show the effectiveness comparison results utilizing another widely used security dataset NSL-KDD. The results are shown in terms of accuracy (\%), precision, recall, and F-score, for different machine learning classifier based anomaly detection models considering binary classification. The results in Table \ref{NSL-KDD-features-comparison-anomaly} are shown for the top five selected features according to their correlation scores and ranking. If we observe the results in Table \ref{NSL-KDD-features-comparison-anomaly}, we can see that almost all the security models give significant results (accuracy 99\%) with the selected top 5 features. Thus, we can conclude that machine learning-based security models are highly dependent on the quality and characteristics of the data, and may give different results for different datasets.

\subsection{Effectiveness Comparison for Detecting Multi-Attacks}
To show the effectiveness of the security models based on machine learning classifiers, Table \ref{UNSW15-features-comparison-attacks} shows the effectiveness comparison results in terms of accuracy (\%) for different machine learning classifier based attacks detection models considering multi-class classification. The results in Table \ref{UNSW15-features-comparison-attacks} are shown by varying the number of selected features such as 42, 31, 24, and 17 utilizing the dataset UNSW-NB15. These features are selected similarly, i.e., according to their correlation scores and ranking considering a particular threshold. If we observe the results in Table \ref{UNSW15-features-comparison-attacks}, we can see that various machine learning security models for detecting multi-attacks have also an impact on the number of selected features. As higher accuracy results with a minimum number of features represent the effectiveness of the security models, the RF model is effective with the accuracy (83\%) when the top 31 features are selected to build the model. Similarly, XGBoost, DT, and SVM security models also give higher accuracy (81\%), (81\%), and (79\%), when the top 31 features are selected to build the corresponding models. Several security models such as NB, LDA, KNN, and AdaBoost show their significant results considering the top 24 features, while the LR model shows significant results considering all the 42 features. Overall, in addition to RF (accuracy 83\%), DT (accuracy 81\%), and XGBoost (accuracy 81\%) also give significant results for detecting multi-attacks. 

\begin{table}[H]
	\centering
	\tiny
	\caption{Effectiveness comparison results in terms of accuracy (\%) for different machine learning classifier based multi-attacks detection models utilizing the dataset UNSW-NB15.}
	\label{UNSW15-features-comparison-attacks}
	\begin{tabular}{ccccc} 
		
		\textbf{Model} & \textbf{Features (42)} &  \textbf{Features (31)}  &  \textbf{Features (24)}  &  \textbf{Features (17)} \\  
		
		NB &	43 &	43 &	\textbf{44} &	42 \\
		LDA &	67 &	67 &	\textbf{67} &	65 \\
		KNN	& 76 &	77 &	\textbf{77} &	73 \\
		XGBoost &	81 &	\textbf{81} &	80 &	76 \\
		DT	& 81 &	\textbf{81} &	80 &	77 \\
		RF	& 83 &	\textbf{83} &	82 &	80 \\
		SVM	& 79 &	\textbf{79} &	78 &	74 \\
		AdaBoost & 	51 &	31 &	\textbf{62} &	57 \\
		SGD	& 71 &	\textbf{72} &	70 &	63 \\
		LR	& \textbf{76} &	75 &	74 &	71 \\
	\end{tabular}
\end{table}

\begin{figure*}[htbp!]
	\centering
	\begin{subfigure}[b]{.49\textwidth}
		\includegraphics[width=\textwidth]{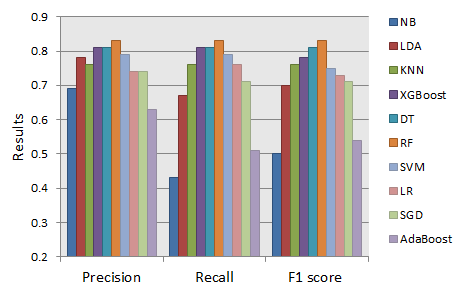}
		\caption{Multi-attacks detection with all 42 features}
		\label{fig:UNSW15-42-features-attacks}
	\end{subfigure}
	\begin{subfigure}[b]{.49\textwidth}
		\includegraphics[width=\textwidth]{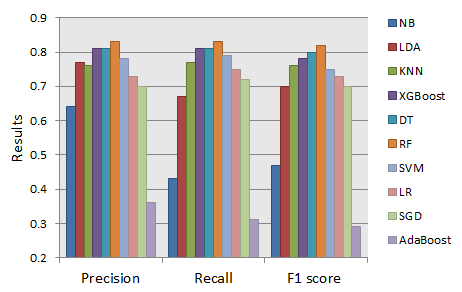}
		\caption{Multi-attacks detection with top 31 features}
		\label{fig:UNSW15-31-features-attacks}
	\end{subfigure}
	\begin{subfigure}[b]{.49\textwidth}
		\includegraphics[width=\textwidth]{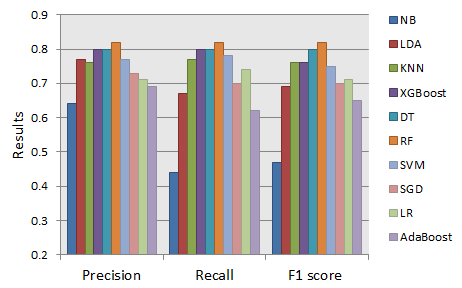}
		\caption{Multi-attacks detection with top 24 features}
		\label{fig:UNSW15-24-features-attacks}
	\end{subfigure}
	\begin{subfigure}[b]{.49\textwidth}
		\includegraphics[width=\textwidth]{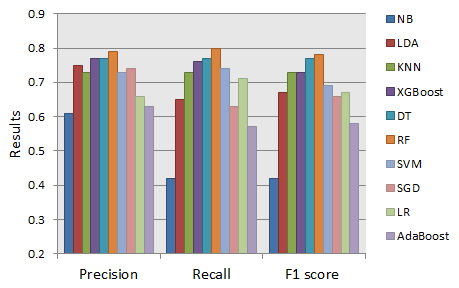}
		\caption{Multi-attacks detection with top 17 features}
		\label{fig:UNSW15-17-features-attacks}
	\end{subfigure}
	\caption{Effectiveness comparison results in terms of precision, recall, and F1 score for different machine learning classifier based multi-attacks detection models utilizing the dataset UNSW-NB15.}
	\label{fig:UNSW15-features-comparison-attacks}
\end{figure*}

In addition to Table \ref{UNSW15-features-comparison-attacks}, Figure \ref{fig:UNSW15-features-comparison-attacks} also shows the relative comparison of various security models based on machine learning classifiers for detecting multi-attacks in terms of precision, recall, and F1 score utilizing the dataset UNSW-NB15. For each security model, we use the same train and testing data to calculate these metrics for fair evaluation.

If we observe Figure \ref{fig:UNSW15-features-comparison-attacks}, we find that tree-based classification models also provide higher prediction results in terms of accuracy, recall, and F1 score, for multi-attack detection than other security models. In particular, the security model based on RF (Random Forest) generating multiple decision trees gives the prediction results with the highest accuracy, recall, and F1 score values, shown in Figure \ref{fig:UNSW15-features-comparison-attacks}. The interesting finding is that like the anomaly detection model, the RF model gives similar results with the features of 42, 31, and 24, and a comparatively lower result with the feature 17 for multi-attack detection. The reason for decreasing the result is that it losses significant information while reducing the features. Thus, the RF model with the top 24 security features is taken into account as an effective model considering both the accuracy and model complexity. Overall, we can conclude that the RF model gives better results in detecting multi-attacks based on the selected security features. The reason is that the random forest model generates a set of logic rules for the attacks based on the selected security features considering several decision trees generated in the forest, and provide an outcome based on the majority voting of these trees.

\begin{table}[H]
	\centering
	\tiny
	\caption{Effectiveness comparison results in terms of accuracy (\%), precision, recall, and F1 score for different machine learning classifier based multi-attacks detection models utilizing the dataset NSL-KDD.}
	\label{NSL-KDD-features-comparison-attacks}
	\begin{tabular}{ccccc} 
		
		\textbf{Model} & \textbf{Accuracy (\%)} &  \textbf{Precision}  &  \textbf{Recall}  &  \textbf{F1 Score} \\  
		
		NB &	91 &	0.97 &	0.91 &	0.93 \\
		LDA &	96 &	0.98 &	0.96 &	0.97 \\
		KNN	& 99 &	0.99 &	0.99 &	0.99 \\
		XGBoost &	99 &	0.99 &	0.99 &	0.99 \\
		DT	& 99 &	0.99 &	0.99 &	0.99 \\
		RF	& 99 &	0.99 &	0.99 &	0.99 \\
		SVM	& 99 &	0.98 &	0.99 &	0.99 \\
		AdaBoost & 	91 &	0.91 &	0.91 &	0.90 \\
		SGD	& 98 &	0.98 &	0.98 &	0.98 \\
		LR	& 98 &	0.98 &	0.98 &	0.98 \\
	\end{tabular}
\end{table}

In Table \ref{NSL-KDD-features-comparison-attacks}, we also show the effectiveness comparison results utilizing another widely used security dataset NSL-KDD. The results are shown in terms of accuracy (\%), precision, recall, and F-score, for different machine learning classifier based multi-attacks detection models considering multi-class classification. The results in Table \ref{NSL-KDD-features-comparison-attacks} are shown for the top five selected features according to their correlation scores and ranking. If we observe the results in Table \ref{NSL-KDD-features-comparison-attacks}, we can see that most of the security models such as KNN, XGBoost, DT, RF, and SVM give the highest results (accuracy 99\%) with the selected top 5 features. The other models also give significant results. Based on the results discussed above, we can conclude that machine learning-based security models are highly dependent on the quality and characteristics of the data, and may give different results for different datasets.

\subsection{Effectiveness Analysis for Neural Network-based Security Model}
\begin{figure*}[htbp!]
	\centering
	\begin{subfigure}[b]{.24\textwidth}
		\includegraphics[width=\textwidth]{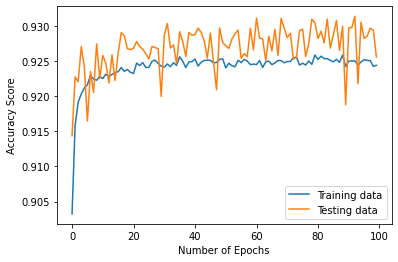}
		\caption{Accuracy score with all 42 features.}
		\label{fig:UNSW15-42-features-anomaly-accuracy-DNN}
	\end{subfigure}
	\begin{subfigure}[b]{.24\textwidth}
		\includegraphics[width=\textwidth]{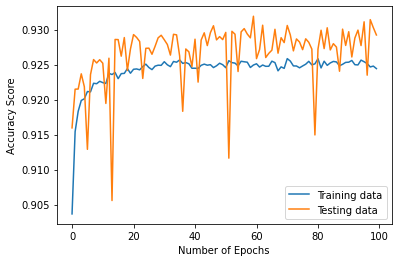}
		\caption{Accuracy score with the top 31 features.}
		\label{fig:UNSW15-31-features-anomaly-accuracy-DNN}
	\end{subfigure}
	\begin{subfigure}[b]{.24\textwidth}
		\includegraphics[width=\textwidth]{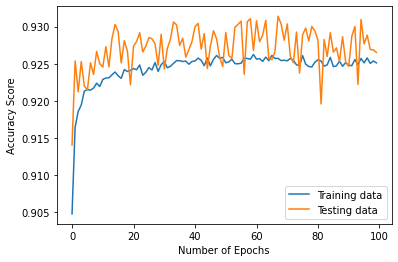}
		\caption{Accuracy score with the top 24 features.}
		\label{fig:UNSW15-24-features-anomaly-accuracy-DNN}
	\end{subfigure}
	\begin{subfigure}[b]{.24\textwidth}
		\includegraphics[width=\textwidth]{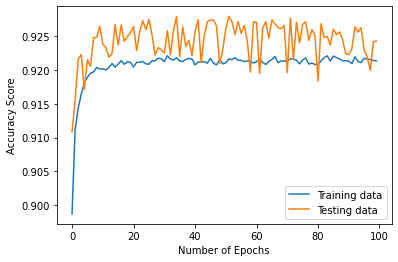}
		\caption{Accuracy score with the top 17 features.}
		\label{fig:UNSW15-17-features-anomaly-accuracy-DNN}
	\end{subfigure}
	
	\begin{subfigure}[b]{.24\textwidth}
		\includegraphics[width=\textwidth]{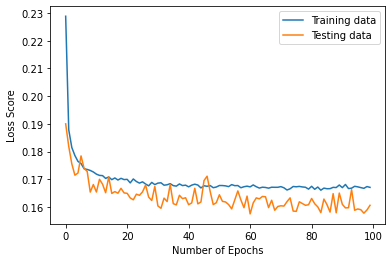}
		\caption{Loss score with all 42 features.}
		\label{fig:UNSW15-42-features-anomaly-loss-DNN}
	\end{subfigure}
	\begin{subfigure}[b]{.24\textwidth}
		\includegraphics[width=\textwidth]{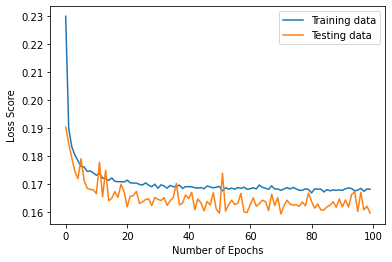}
		\caption{Loss score with the top 31 features.}
		\label{fig:UNSW15-31-features-anomaly-loss-DNN}
	\end{subfigure}
	\begin{subfigure}[b]{.24\textwidth}
		\includegraphics[width=\textwidth]{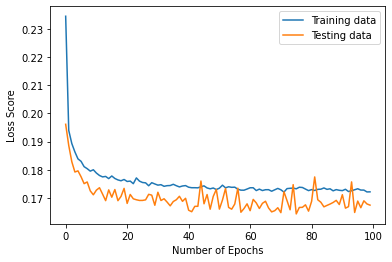}
		\caption{Loss score with the top 24 features.}
		\label{fig:UNSW15-24-features-anomaly-loss-DNN}
	\end{subfigure}
	\begin{subfigure}[b]{.24\textwidth}
		\includegraphics[width=\textwidth]{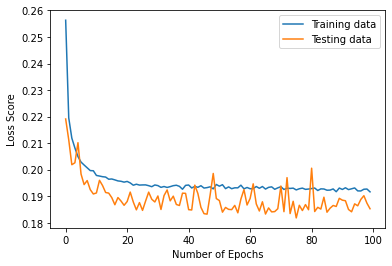}
		\caption{Loss score with the top 17 features.}
		\label{fig:UNSW15-17-features-anomaly-loss-DNN}
	\end{subfigure}
	\caption{Calculated outcome in terms of accuracy and loss score of the deep neural network based security model for detecting anomalies utilizing the dataset UNSW-NB15.}
	\label{fig:UNSW15-anomaly-DNN}
\end{figure*}

\begin{figure*}[h!]
	\centering
	\begin{subfigure}[b]{.24\textwidth}
		\includegraphics[width=\textwidth]{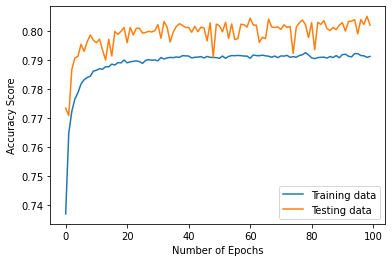}
		\caption{Accuracy score with all 42 features.}
		\label{fig:UNSW15-42-features-attacks-accuracy-DNN}
	\end{subfigure}
	\begin{subfigure}[b]{.24\textwidth}
		\includegraphics[width=\textwidth]{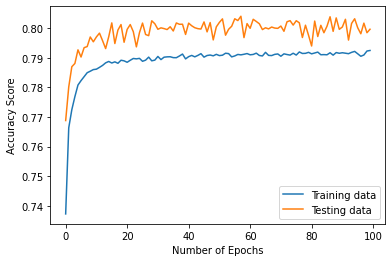}
		\caption{Accuracy score with the top 31 features.}
		\label{fig:UNSW15-31-features-attacks-accuracy-DNN}
	\end{subfigure}
	\begin{subfigure}[b]{.24\textwidth}
		\includegraphics[width=\textwidth]{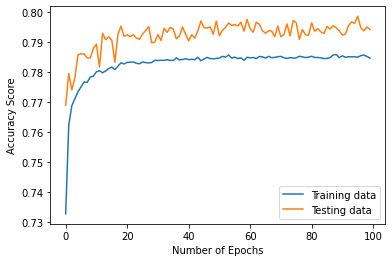}
		\caption{Accuracy score with the top 24 features.}
		\label{fig:UNSW15-24-features-attacks-accuracy-DNN}
	\end{subfigure}
	\begin{subfigure}[b]{.24\textwidth}
		\includegraphics[width=\textwidth]{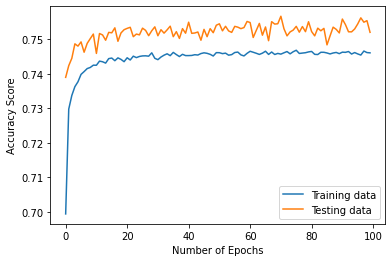}
		\caption{Accuracy score with the top 17 features.}
		\label{fig:UNSW15-17-features-attacks-accuracy-DNN}
	\end{subfigure}
	
	\begin{subfigure}[b]{.24\textwidth}
		\includegraphics[width=\textwidth]{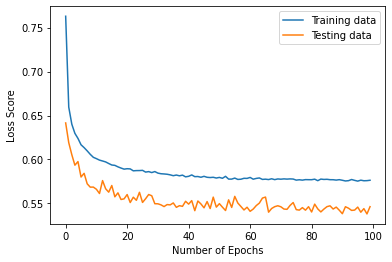}
		\caption{Loss score with all 42 features.}
		\label{fig:UNSW15-42-features-attacks-loss-DNN}
	\end{subfigure}
	\begin{subfigure}[b]{.24\textwidth}
		\includegraphics[width=\textwidth]{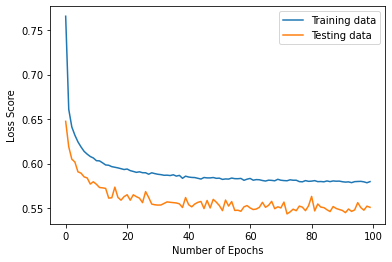}
		\caption{Loss score with the top 31 features.}
		\label{fig:UNSW15-31-features-attacks-loss-DNN}
	\end{subfigure}
	\begin{subfigure}[b]{.24\textwidth}
		\includegraphics[width=\textwidth]{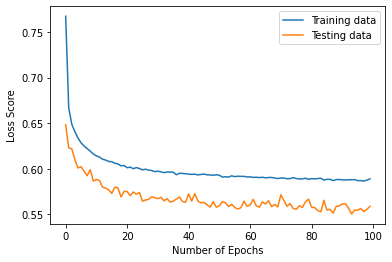}
		\caption{Loss score with the top 24 features.}
		\label{fig:UNSW15-24-features-attacks-loss-DNN}
	\end{subfigure}
	\begin{subfigure}[b]{.24\textwidth}
		\includegraphics[width=\textwidth]{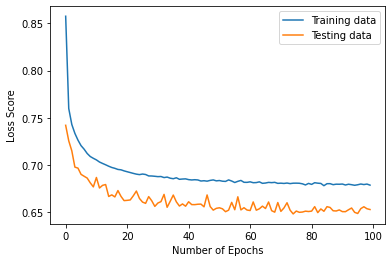}
		\caption{Loss score with the top 17 features.}
		\label{fig:UNSW15-17-features-attacks-loss-DNN}
	\end{subfigure}
	\caption{Calculated outcome in terms of accuracy and loss score of the neural network based security model for detecting multi-attacks utilizing the dataset UNSW-NB15.}
	\label{fig:UNSW15-attacks-DNN}
\end{figure*}

To show the model effectiveness based on artificial neural network, Figure \ref{fig:UNSW15-anomaly-DNN} shows the calculated outcome in terms of model accuracy and loss score for detecting anomalies considering binary classification. The results in Figure \ref{fig:UNSW15-anomaly-DNN} are shown by varying the number of selected features such as 42, 31, 24, and 17 utilizing the dataset UNSW-NB15. The features are selected similarly, according to their correlation scores and ranking, shown in Table \ref{feature-score} considering a particular threshold mentioned above. Similarly, for multi-attacks classification, Figure \ref{fig:UNSW15-attacks-DNN} shows the calculated outcome in terms of model accuracy and loss score considering multi-class classification according to our goal. For each neural network-based security model, we use the same train and testing data for fair evaluation and comparison.

If we observe the results in Figure \ref{fig:UNSW15-anomaly-DNN} and Figure \ref{fig:UNSW15-attacks-DNN}, we can see that a neural network-based security model with a variable number of selected features can detect both the anomalies and multi-attacks. Similar to classic machine learning classification models, discussed above, we get higher accuracy results in anomaly detection using the neural network-based security model. According to Figure \ref{fig:UNSW15-anomaly-DNN}, the model with the top 24 features gives the results of 92\% accuracy with a loss of 0.1681, which is significant in terms of accuracy and complexity, comparing with other models with different number of features, shown in Figure  \ref{fig:UNSW15-anomaly-DNN}. Thus model with the top 24 features can be selected as an effective security model that gives significant accuracy with a reduced number of features for detecting anomalies. Similarly, a model with the top 24 features can also be selected as an effective model for detecting multi-attacks, shown in Figure \ref{fig:UNSW15-attacks-DNN}.

\begin{figure*}[htbp!]
	\centering
	\begin{subfigure}[b]{.24\textwidth}
		\includegraphics[width=\textwidth]{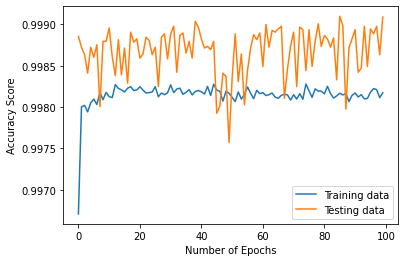}
		\caption{Accuracy score with all 42 features.}
		\label{fig:KDD-42-features-anomaly-accuracy-DNN}
	\end{subfigure}
	\begin{subfigure}[b]{.24\textwidth}
		\includegraphics[width=\textwidth]{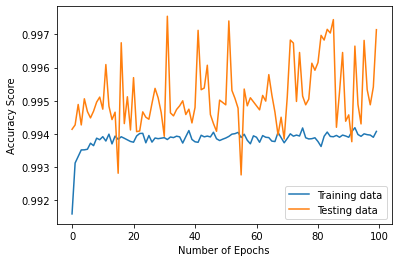}
		\caption{Accuracy score with the top 27 features.}
		\label{fig:KDD-27-features-anomaly-accuracy-DNN}
	\end{subfigure}
	\begin{subfigure}[b]{.24\textwidth}
		\includegraphics[width=\textwidth]{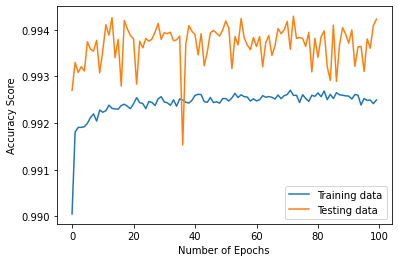}
		\caption{Accuracy score with the top 18 features.}
		\label{fig:KDD-18-features-anomaly-accuracy-DNN}
	\end{subfigure}
	\begin{subfigure}[b]{.24\textwidth}
		\includegraphics[width=\textwidth]{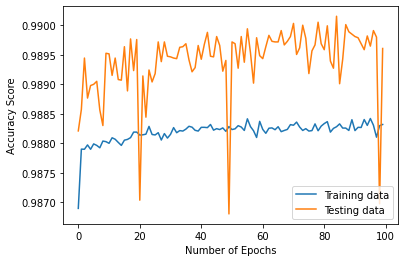}
		\caption{Accuracy score with the top 5 features.}
		\label{fig:KDD-5-features-anomaly-accuracy-DNN}
	\end{subfigure}
	
	\begin{subfigure}[b]{.24\textwidth}
		\includegraphics[width=\textwidth]{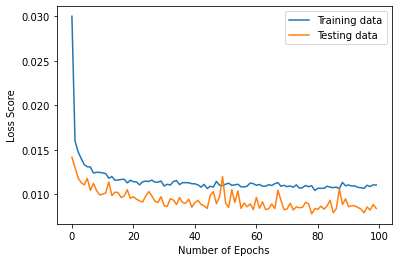}
		\caption{Loss score with all 42 features.}
		\label{fig:KDD-42-features-anomaly-loss-DNN}
	\end{subfigure}
	\begin{subfigure}[b]{.24\textwidth}
		\includegraphics[width=\textwidth]{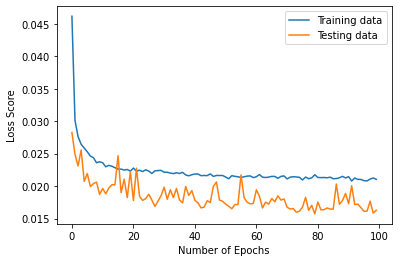}
		\caption{Loss score with the top 27 features.}
		\label{fig:KDD-27-features-anomaly-loss-DNN}
	\end{subfigure}
	\begin{subfigure}[b]{.24\textwidth}
		\includegraphics[width=\textwidth]{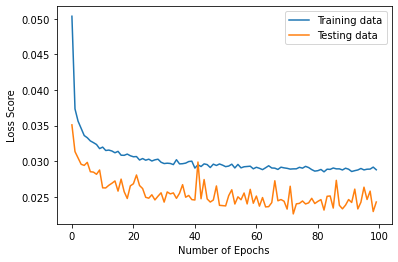}
		\caption{Loss score with the top 18 features.}
		\label{fig:KDD-18-features-anomaly-loss-DNN}
	\end{subfigure}
	\begin{subfigure}[b]{.24\textwidth}
		\includegraphics[width=\textwidth]{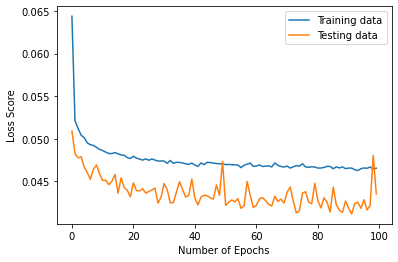}
		\caption{Loss score with the top 5 features.}
		\label{fig:KDD-5-features-anomaly-loss-DNN}
	\end{subfigure}
	\caption{Calculated outcome in terms of accuracy and loss score of the deep neural network based security model for detecting anomalies utilizing the dataset NSL-KDD.}
	\label{fig:KDD-anomaly-DNN}
\end{figure*}

\begin{figure*}[h!]
	\centering
	\begin{subfigure}[b]{.24\textwidth}
		\includegraphics[width=\textwidth]{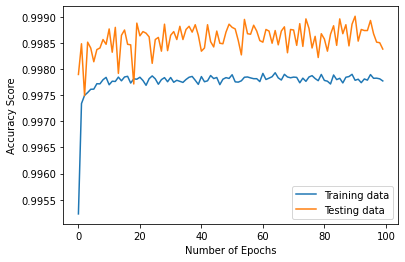}
		\caption{Accuracy score with all 42 features.}
		\label{fig:KDD-42-features-attacks-accuracy-DNN}
	\end{subfigure}
	\begin{subfigure}[b]{.24\textwidth}
		\includegraphics[width=\textwidth]{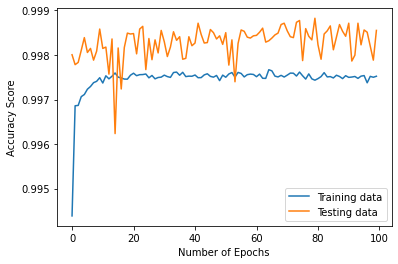}
		\caption{Accuracy score with the top 27 features.}
		\label{fig:KDD-27-features-attacks-accuracy-DNN}
	\end{subfigure}
	\begin{subfigure}[b]{.24\textwidth}
		\includegraphics[width=\textwidth]{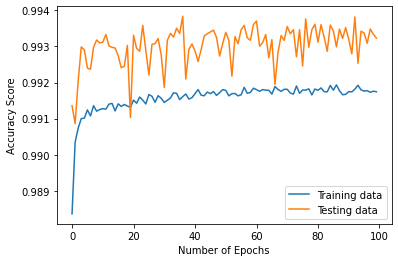}
		\caption{Accuracy score with the top 18 features.}
		\label{fig:KDD-18-features-attacks-accuracy-DNN}
	\end{subfigure}
	\begin{subfigure}[b]{.24\textwidth}
		\includegraphics[width=\textwidth]{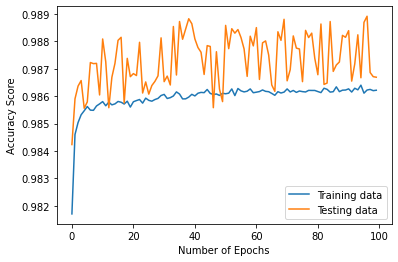}
		\caption{Accuracy score with the top 5 features.}
		\label{fig:KDD-5-features-attacks-accuracy-DNN}
	\end{subfigure}
	
	\begin{subfigure}[b]{.24\textwidth}
		\includegraphics[width=\textwidth]{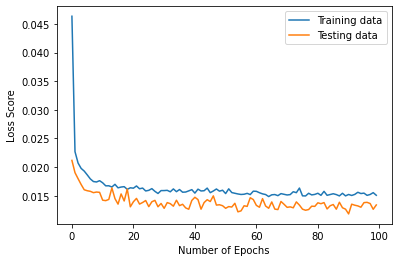}
		\caption{Loss score with all 42 features.}
		\label{fig:KDD-42-features-attacks-loss-DNN}
	\end{subfigure}
	\begin{subfigure}[b]{.24\textwidth}
		\includegraphics[width=\textwidth]{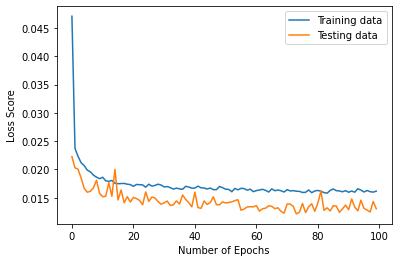}
		\caption{Loss score with the top 27 features.}
		\label{fig:KDD-27-features-attacks-loss-DNN}
	\end{subfigure}
	\begin{subfigure}[b]{.24\textwidth}
		\includegraphics[width=\textwidth]{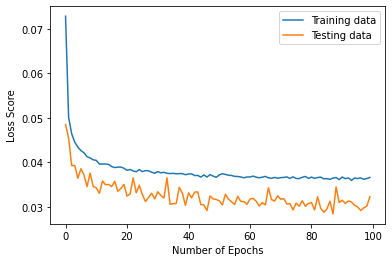}
		\caption{Loss score with the top 18 features.}
		\label{fig:KDD-18-features-attacks-loss-DNN}
	\end{subfigure}
	\begin{subfigure}[b]{.24\textwidth}
		\includegraphics[width=\textwidth]{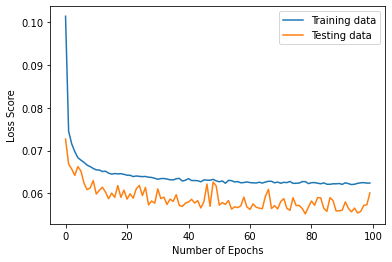}
		\caption{Loss score with the top 5 features.}
		\label{fig:KDD-5-features-attacks-loss-DNN}
	\end{subfigure}
	\caption{Calculated outcome in terms of accuracy and loss score of the deep neural network based security model for detecting multi-attacks utilizing the dataset NSL-KDD.}
	\label{fig:KDD-attacks-DNN}
\end{figure*}

Besides, Figure \ref{fig:KDD-anomaly-DNN} shows the calculated outcome in terms of model accuracy and loss score for detecting anomalies considering binary classification utilizing another widely used dataset NSL-KDD. The results in Figure \ref{fig:KDD-anomaly-DNN} are shown by varying the number of selected features such as 42, 27, 18, and 5 utilizing the dataset NSL-KDD. These are selected according to their correlation scores and ranking considering a particular threshold as well. Similarly, for multi-attacks classification, Figure \ref{fig:KDD-attacks-DNN} shows the calculated outcome in terms of model accuracy and loss score considering multi-class classification. According to Figure \ref{fig:KDD-anomaly-DNN}, the model with the top 18 features gives the results of 99\% accuracy with a loss of 0.0243, which is significant in terms of accuracy and complexity, comparing with other models with different number of features, shown in Figure  \ref{fig:KDD-anomaly-DNN}. Thus model with the top 18 features can be selected as an effective security model that gives significant accuracy with a reduced number of features for detecting anomalies. Similarly, a model with the top 27 features can also be selected as an effective model for detecting multi-attacks, shown in Figure \ref{fig:KDD-attacks-DNN}.

\section{Discussion}
\label{Discussion}
Overall, our CyberLearning model based on machine learning approaches is fully security data-oriented that reflects the data patterns related to the security incidents, e.g., cyber-anomalies and attacks, according to our goal. The model can effectively detect anomalies and different types of attacks, such as DoS, Backdoor, Worms, etc, where the popular machine learning classification techniques including artificial neural network models are employed. The experimental analysis on the UNSW-NB15 \cite{moustafa2015unsw} and NSL-KDD \cite{tavallaee2009detailed} datasets, have shown the effectiveness of the resultant security models according to their learning capabilities in various situations, as discussed in the earlier section. 

According to the experimental analysis discussed in Section \ref{Evaluation}, we can say that different machine learning-based security models perform differently for detecting cyber-anomalies, or multi-attacks utilizing the training security data. The significance of the security features greatly impact both the binary classification model while detecting anomalies for unknown attacks, as well as the multi-class classification model while detecting several known classes mentioned above. For instance, according to experimental results shown in Table \ref{feature-score}, the feature $sttl$ has the highest correlation score of $0.624082$ and thus selected as the highly significant feature, whereas another feature $ackdat$ has a lower score of $0.000817$ that is closer to the value $0$ for the dataset UNSW-NB15 \cite{moustafa2015unsw}, and thus can be considered as the less significant feature for modeling. A set of highly significant security features reducing the insignificant or irrelevant features can help to make the security model lightweight and more applicable. For instance, the NB security model gives higher accuracy (85\%) when the top 24 features are selected for detecting cyber-anomalies, rather than considering all 42 features as shown in Table \ref{UNSW15-features-comparison-anomaly}. Overall, the security models for detecting anomalies and attacks, based on various learning algorithms are also affected by the variations in the significance of the security features, as discussed briefly in Section \ref{Evaluation}. 

Besides, a robust classification model is essential to the design of an intelligent intrusion detection system. The reason is that the performance of all machine learning classification techniques are not identical in the real world scenario, depending on their learning capabilities from the security data. As shown in Table \ref{UNSW15-features-comparison-anomaly}, and Figure \ref{fig:UNSW15-features-comparison-anomaly}, the RF-based security model generating multiple decision trees, give higher prediction results for detecting anomalies than other security models, in terms of accuracy, precision, recall, and F1 score. Several other models such as DT, XGBoost, SVM, KNN, AdaBoost also give significant outcome based on the selected features. According to the results, shown in Table \ref{NSL-KDD-features-comparison-anomaly}, the RF model also gives higher accuracy for detecting anomalies. As shown in the Table \ref{UNSW15-features-comparison-attacks}, Figure \ref{fig:UNSW15-features-comparison-attacks}, the RF-based security model also gives higher prediction results for detecting multi-attacks than other security models, in terms of accuracy, precision, recall, and F1 score. Several other models such as DT and XGBoost also give significant outcome based on the selected features while detecting multi-attacks. According to the results, shown in Table \ref{NSL-KDD-features-comparison-anomaly}, the RF model also gives higher accuracy for detecting multi-attacks. 

The model performance for detecting multi-attacks may differ with the anomaly detection mentioned above, even for the same learning technique. For instance, the accuracy of the RF security model for detecting anomalies is 95\%, and 83\% for multi-attacks detection for the dataset UNSW-NB15 \cite{moustafa2015unsw}. Similarly, it achieves 99\% for both cases for the dataset NSL-KDD \cite{tavallaee2009detailed}. Thus, we can say that the effectiveness of a learning-based security model may vary depending on the security features and the data characteristics. Overall, we can conclude that RF (Random Forest) based security model is more effective for detecting anomalies and multi-attacks. The reason is that the random forest model has the learning capabilities considering several decision trees that generate a set of logic rules based on the selected security features. Thus, the model gives higher prediction results in terms of accuracy, precision, recall, and F1 score.

A real-life cybersecurity application is the actual platform to use the CyberLearning model that typically examines the behavior of the network, finding the security patterns for profiling the normal behavior, and thus detects the anomalies or associated attacks. Although an ANN model has its hidden layers for computing, it also affects on the significance of the features. For instance, an ANN model with the selected security features gives significant accuracy for detecting anomalies and multi-attacks, as discussed briefly in Section \ref{Evaluation}. Although we use the security datasets UNSW-NB15 \cite{moustafa2015unsw} and NSL-KDD \cite{tavallaee2009detailed} while building the security model, our analysis is also applicable to other application domains in the area of cybersecurity, including IoT security. Several deep learning networks such as Convolutional neural network (CNN), recurrent neural network (RNN), Long Short-Term Memory (LSTM), deep belief network (DBN), or an autoencoder, etc. could be effective while working on a huge number of datasets. Typically deep learning algorithms perform well when the data volumes are large \cite{sarker2020cybersecurity}  \cite{xin2018machine}. In addition, noisy instance analysis \cite{sarker2019machine}, incorporating contextual information \cite{sarker2019context} \cite{sarker2020mobile}, or recency analysis considering recent patterns in data \cite{sarker2019recencyminer}, could be another potential research dimensions in the area. Overall, we believe that our CyberLearning model including a comprehensive experimental analysis opens a promising path for future research in the domain of cybersecurity, while working on machine learning-based security modeling, to make the security model lightweight and more applicable in the area.

\section{Conclusion and Future Work}
\label{Conclusion}
In this paper, we have presented CyberLearning, where we have taken into account a binary classification model for detecting anomalies and a multi-class classification model for various types of cyber-attacks. In our modeling, we have also taken into account the impact of security features, and eventually built a machine learning based effective model with feature selection. While building the security models, we have employed the most popular machine learning classification techniques as well as artificial neural network learning considering multiple hidden layers. Finally, we have examined the effectiveness of these learning-based security models by conducting a range of experiments utilizing the two most popular security datasets, UNSW-NB15 and NSL-KDD. We believe that our empirical analysis and findings can be used as a reference guide in both academia and industry in the area of cybersecurity for effectively building a data-driven security modeling and system based on machine learning techniques. 

To collect more recent security data with higher dimensions in the environment of IoT, and build a data-driven secure system using learning techniques could be a future work.

\bibliographystyle{plain}
\bibliography{Cybersecurity}

\end{document}